\documentstyle[sprocl,psfig]{article}

\def\be{\begin{equation}}
\def\ee{\end{equation}}
\def\bea{\begin{eqnarray}}
\def\eea{\end{eqnarray}}
\begin{document}

\title{DYNAMICAL BREAKING OF CPT AND BARYOGENESIS}

\author{ROBERT H. BRANDENBERGER}

\address{Physics Department, Brown University\\ Providence, RI 02912, USA\\E-mail: rhb@het.brown.edu} 


\maketitle
\abstracts{The asymmetry between matter and antimatter in the 
Universe indicates that there was a period in the very early Universe when the
CPT symmetry was broken. The conservative interpretation is that this CPT
breaking was dynamical, induced by the distinguished direction of time
stemming from the cosmological expansion. Here, I review the role which 
topological defect networks may play in baryogenesis and how this relates
to dynamical CPT symmetry breaking. Particular attention is placed on
defect-mediated electroweak baryogenesis. A recent suggestion of defect-mediated QCD scale baryogenesis is also mentioned.}

\noindent{BROWN-HET-1163, January 1999 \footnote{Invited talk at CPT-98,
Bloomington, Indiana, Nov. 6 - 8 1998, to be
publ. in the proceedings (World Scientific, Singapore, 1999).}}

\section{Introduction}

The mass in the visible Universe appears to be made up exclusively of matter. There is no evidence for stable antimatter up to close to the present Hubble radius \cite{Cohen}. Based on the observed matter energy density (from which the number density $n_B$ of baryons can be determined) and on the measured temperature of the cosmic microwave background (which yields the entropy density $s$), it follows that the baryon to entropy ratio $n_B / s$ is 
\begin{equation}
{{n_B} \over s} \, \sim \, 10^{-10} \, .
\label{ratio}
\end{equation}
A second argument in favor of this value of $n_B / s$ comes from the theory of big bang nucleosynthesis \cite{BBN}. The predicted and observed abundances of the light elements agree precisely if $n_B / s$ is in the range given by (\ref{ratio}). The goal of the theory of baryogenesis is to explain the origin of (\ref{ratio}) starting with symmetric initial conditions at very early times.

Sakharov \cite{Sakharov} realized that in order to obtain a model of baryogenesis, three criteria must be satisfied:\hfill\break
1. $n_B$ violating processes must exist;\hfill\break
2. these processes must involve C and CP violation;\hfil\break
3. they must occur out of thermal equilibrium.\hfil\break
Another way to state these criteria is that, in addition to the existence of baryon number violating processes, there needs to be a period in the early Universe in which the CPT symmetry is broken. In an expanding Universe, this condition is not hard to achieve since the expansion determines a preferred direction of time.

The first theory of baryogenesis (there are several good recent reviews \cite{GUTBG} on this topic) was in the context of Grand Unified Theories, theories in which baryon number violating processes occur at the perturbative level since there are particles (the superheavy Higgs and gauge particles $X$ and $A_{\mu}$) which couple baryons and leptons. Baryons are generated at a temperature $T_{out} \sim T_{GUT} \sim 10^{16}{\rm GeV}$ by the out-of-equilibrium decay of the superheavy $X$ and $A_{\mu}$ particles. These particles were in thermal equilibrium for $T \gg T_{GUT}$ but fall out of equilibrium at a temperature $T_{out}$ close to the GUT symmetry breaking scale $T_{GUT}$ as the Universe expands and cools.

Obviously, GUT baryogenesis makes use of new physics beyond the standard model. It also requires a new source of CP violation (perturbative CP violation in the standard model sector is too weak to account for the observed value of $n_B / s$), but such new CP violation is rather naturally present in the extended Higgs sector of a GUT model.

A potentially fatal problem for GUT baryogenesis was pointed out by Kuzmin, Rubakov and Shaposhnikov \cite{KRS}: there are nonperturbative processes in the standard model which violate baryon number and are unsuppressed for $T \gg T_{EW}$, where $T_{EW}$ is the electroweak symmetry breaking scale, and hence can erase any primordial baryon asymmetry generated at $T \gg T_{EW}$, for example $T = T_{GUT}$. One way to protect GUT baryogenesis from this washout is to generate during the GUT phase transition an asymmetry in a quantum number like $B - L$ (where $B$ and $L$ denote baryon and lepton number, respectively) which is not violated by nonperturbative electroweak processes.

The nonperturbative baryon number violating processes in the electroweak theory are related to the nontrivial gauge vacuum structure \cite{JR,tHooft}. The configuration $A_{\mu} = 0$ is not the only vacuum state. There are energetically degenerate states with nontrivial gauge field configurations $A_{\mu} \neq 0$. A gauge-invariant way to distinguish between these states is in terms of a topological invariant, the Chern-Simons number $N_{CS}$. The transitions between configurations of different $N_{CS}$ are called {\it sphaleron} transitions \cite{Manton}. They are exponentially suppressed at zero temperature $T = 0$. However, at temperatures $T \gg T_{EW}$, they are unsuppressed. In a theory in which $N_f$ fermion SU(2) doublets couple to the gauge fields, there is a change in baryon number $\Delta N_B$ associated with a sphaleron transition:
\begin{equation}
\Delta N_B \, = \, N_f \Delta N_{CS} \, .
\end{equation}
Hence, for $T \gg T_{EW}$, baryon number violating processes are in equilibrium. Note, however, that sphalerons preserve $B - L$. 

An alternative to trying to protect a primordial matter asymmetry generated at some temperature $T \gg T_{EW}$ from sphaleron washout is to make use of out-of-equilibrium sphaleron processes at $T \ll T_{EW}$ to re-generate a new baryon number below the electroweak phase transition. This is the goal of electroweak baryogenesis. Following early work by Shaposhnikov \cite{Shap} and Arnold and McLerran \cite{AMcL}, concrete models of electroweak baryogenesis were suggested by Turok and Zadrozny \cite{Turok} and by Cohen, Kaplan and Nelson \cite{CKN}. These mechanisms were based on sphaleron processes inside or in the vicinity of bubble walls nucleated at the electroweak phase transition. These mechanisms require the electroweak phase transition to be strongly first order and nucleation-driven. In this case, below the critical temperature $T_{EW}$, bubbles of the low temperature vacuum are nucleated in the surrounding sea of the false (i.e. high temperature) vacuum and then expand until they percolate. Detailed studies (see recent review articles \cite{EWBGrev} for details) indicate that physics beyond the standard model is needed in order to implement the mechanism, specifically in order for the phase transition to be strongly first order and to obtain sufficient CP violation.

In this light, defect-mediated electroweak baryogenesis \cite{BD,BDPT} may be a promising alternative, since many theories beyond the standard model predict topological defects. In this case, the baryogenesis mechanism involves sphaleron processes inside the topological defects. In the following sections, I will review the defect-mediated electroweak baryogenesis mechanism and discuss how the dynamical breaking of CPT symmetry in defect networks leads to a nonvanishing net baryon number. These sections are based on \cite{BDPT} and \cite{PTDB}, respectively. In Section 4 I will mention recent ideas \cite{BHZ} on QCD-scale ``baryogenesis", a charge separation mechanism which also makes crucial use of the effective T violation in the defect dynamics in an expanding Universe.
 
\section{Defect-Mediated Electroweak Baryogenesis}

Before discussing the role of defects in electroweak baryogenesis I will review the main points of the ``standard" or ``first-order" mechanism \cite{Turok,CKN}. It is based on two key assumptions:  
\begin{enumerate}
\item{} The electroweak phase transition is first order.
\item{} The transition is nucleation-driven (rather than fluctuation-driven, see e.g. the article by Goldenfeld \cite{Goldenfeld} for critical comments on transition dynamics from the point of view of condensed matter physics).
\end{enumerate}
\noindent If these assumptions are satisfied, then the electroweak phase transition proceeds by the nucleation of bubbles of the low temperature vacuum in a surrounding sea of the high temperature, symmetric vacuum. Inside the bubbles, the electroweak symmetry is broken and sphalerons are suppressed, outside the bubbles the symmetry is restored and the sphaleron rate is not suppressed. The bubbles are nucleated with microscopic radius and then expand monotonically until they percolate.

Let us briefly consider the way in which the Sakharov criteria are satisfied: The standard electroweak theory contains C and CP violating interactions which couple to the fields excited in the bubble walls (2nd criterium). The bubbles are out of equilibrium field configurations (3rd condition). Baryogenesis occurs via sphaleron processes near the bubble walls (1st criterium). Note that the bubble dynamics (expansion into the false vacuum) represents the effective dynamical breaking of CPT.

The master equation for electroweak baryogenesis is
\begin{equation}
{{dn_B} \over {dt}} \, = \, - 3 \Gamma \mu \, ,
\label{master}
\end{equation}
where
\begin{equation}
\Gamma \, = \, \kappa (\alpha_w T)^4 
\label{rate}
\end{equation}
is the sphaleron rate in the false vacuum \cite{sphrate} ($\alpha_w$ is the electroweak fine structure constant and $\kappa$ is a constant which must be determined in numerical simulations), and $\mu$ is the chemical potential for baryon number which is determined by the interplay between defect dynamics and CP violating interactions of the bubble wall, a complicated issue which is still not fully understood quantitatively. In qualitative terms, fermions scatter off the wall, generating a nonvanishing lepton number in front of the bubble (let us say at point $x$) which yields $\mu(x) \neq 0$ and biases sphaleron processes in front of the wall, yielding $n_B(x) \neq 0$. This value of $n_B(x)$ is then preserved as the wall passes by and the point $x$ becomes part of the true vacuum domain.

The chemical potential $\mu$ is proportional to the constant $\epsilon$ describing the strength of CP violation. In the standard electroweak theory, $\epsilon$ is much too small to account for the observed $n_B / s$. Thus, extra CP violation beyond the standard model is required for successful electroweak baryogenesis. Another reason why physics beyond the standard model is required is that in the context of the basic electroweak model, sphaleron processes are still in equilibrium below $T_{EW}$ if the Higgs mass $m_H$ is larger than $90$GeV, which experimental bounds now indicate must be the case. In addition, for large $m_H$, the phase transition is no longer strongly first order, eliminating the first order baryogenesis mechanism alltogether. Even in the MSSM (the minimal supersymmetric standard model), the window for successful first order electroweak baryogenesis is very small \cite{Cline}.

Hence, extensions of the standard model are required in order to realize baryogenesis at the electroweak scale. Many extensions of the standard model, e.g. theories with additional U(1) gauge symmetries which are broken at or above $T_{EW}$, admit topological defects. In this case, there is an alternative way to implement electroweak baryogenesis which does not make use of bubbles created at a first order transition. Topological defects may replace bubble walls as the out-of-equilibrium field configurations needed to satisfy the third Sakharov criterium.

To be specific, we make the following assumptions:
\begin{enumerate}
\item{} New physics at a scale $\eta > T_{EW}$ generates topological defects.
\item{} The electroweak symmetry is unbroken inside the defects and the defects are sufficiently thick such that sphalerons fit inside them.
\end{enumerate}
Given these assumptions, the scenario for baryogenesis is as follows: At the critical temperature $\eta$, a network of defects is produced by the usual Kibble mechanism \cite{Kibble}. The initial separation of the defects is microscopic, and hence a substantial fraction of space lies inside the defects. As the Universe expands, the defect network coarsens. As long as $T > T_{EW}$, all baryon number violating processes are in equilibrium and hence $n_B = 0$. Once $T$ drops below $T_{EW}$ (more precisely, when $T$ falls below the temperature $T_{out}$ at which sphalerons fall out of equilibrium), then baryon number generation sets in triggered by the defects, in a manner analogous to how bubble walls trigger baryogenesis in the first order mechanism described earlier.

The mechanism can be described with the help of Figure 1. The defect is moving with velocity $v_D$ through the primordial plasma. At the leading edge, a baryon excess of negative sign builds up due to CP violating scatterings from the defect wall. Consider now a point $x$ in space which the defect crosses. When $x$ is hit by the leading defect edge, a value $n_B(x) = - \Delta n_B < 0$ is generated. While $x$ is inside the defect core, this baryon asymmetry relaxes (at least partially) by sphaleron processes. When the trailing edge of the defect passes by, then an asymmetry $\Delta n_B$ of equal magnitude but opposite sign as what is produced at the leading edge is generated. Due to the partial washout in the defect core, the net effect is to produce a positive baryon number density.

\begin{figure}[htbp] 
%\rule{5cm}{0.2mm}\hfill\rule{5cm}{0.2mm}
%\vskip 2.5cm
%\rule{5cm}{0.2mm}\hfill\rule{5cm}{0.2mm}
\centering %\leavemode
\psfig{file=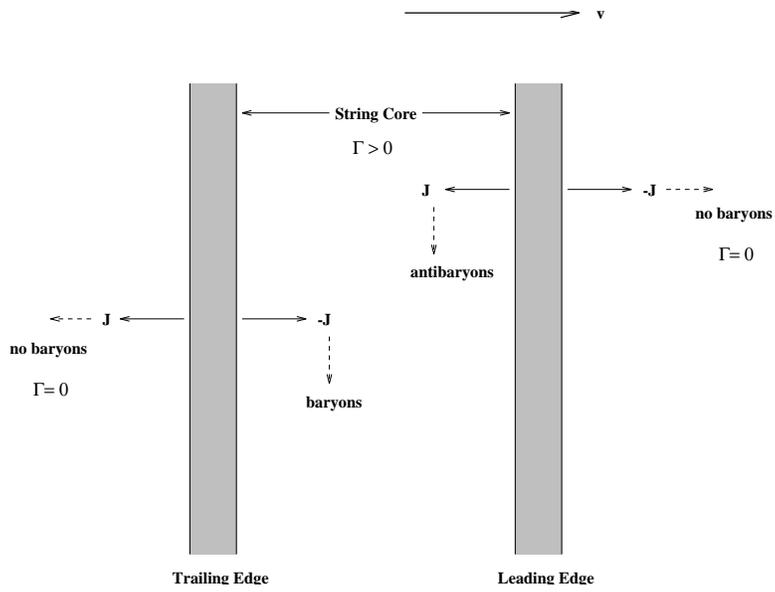,width=4in,height=3in,angle=270,clip=}
\caption{Diagram of a portion of a defect, in this case a cosmic string, moving to the right through the primordial plasma. The differing decays of reflected particles within and outside the defect leads to the generation of a net baryon asymmetry.}
\end{figure}

The same master equation (\ref{master}) as for first order electroweak baryogenesis also applies to defect-mediated electroweak baryogenesis. However, the maximal $n_B / s$ which can be generated from defects is suppressed compared to what could be obtained in successful first order electroweak baryogenesis for several reasons. Most importantly, not all points in space are passed by defects after $T_{EW}$, and hence there is an important geometrical suppression factor $SF$. The value of $SF$ is the fraction of space which will be in a defect core at some time after $t_{EW}$. The value of $SF$ depends sensitively on the type of defect, and on the defect formation scale $\eta$. For non-superconducting cosmic strings \cite{BDPT}
\begin{equation} \label{supp}
SF \, \sim \, \lambda v_D ({{T_{EW}} \over {\eta}})^{3/2} \, ,
\end{equation}
where $\lambda$ is the coupling constant of the string sector which determines the string width and string separation $\xi(t)$ at the time $t_c$ of string formation:
\begin{equation}
\xi(t_c) \, \sim \, \lambda^{-1} \eta^{-1} \, .
\end{equation}
The factor $(T_{EW} / \eta)^{3/2}$ in Equation (\ref{supp}) for the suppression factor comes from the coarsening of the defect network after formation and the resulting growth of $\xi(t)$. Therefore, the fraction of space covered by defects at $T_{EW}$ decreases as the string formation scale $\eta$ increases. For domain walls, there is much less suppression, because of the higher dimensionality of the defects. We find $SF \sim v_D$ \cite{BDPT}. For monopoles, on the other hand, the suppression factor renders defect-mediated baryogenesis completely ineffective.

A further suppression factor comes from having only partial relaxation of $n_B$ inside the defects \cite{Espinosa}. A calculation without taking this latter factor into account yields (for non-superconducting cosmic strings) \cite{BDPT}
\begin{equation}
{{n_B} \over s} \, \sim \, \lambda \kappa \alpha_w^2 g_*^{-1} ({{m_t} \over {T_{EW}}})^2 \epsilon ({{T_{EW}} \over {\eta}})^{3/2}  \, ,
\end{equation}
where $g_*$ gives the number of spin degrees of freedom in the radiation bath,
$\epsilon$ is the CP violating phase, and $m_t$ is the top quark mass. Efficient defect-mediated electroweak baryogenesis thus requires either cosmic strings with $\eta$ close to $T_{EW}$ (plus other optimistic assumptions about the parameters such as $\epsilon \sim 1$ - although according to Cline et al. \cite{Espinosa} even this may not be sufficient), or domain walls (which in turn must decay at late times in order to avoid the domain wall over-abundance problem \cite{DW}).

Defect-mediated electroweak baryogenesis carries the advantage of being independent of the order of the electroweak phase transition and of the Higgs mass. In addition, whereas the efficiency of first-order baryogenesis is exponentially suppressed if $T_{out} < T_{EW}$ (since bubbles are only present
at $T_{EW}$), defect-mediated baryogenesis is only suppressed by a power of
$T_{out} / T_{EW}$ since defects are present at all times after $T_{EW}$. The power is determined by the coarsening dynamics of the defect network.

Note that defect-mediated baryogenesis is not tied to the electroweak scale. Any defects which arise in the early Universe can potentially play a role in baryogenesis, as long as they couple to baryon number violating processes. This applies in particular to defects formed at the GUT scale. GUT defect-mediated baryogenesis \cite{BDH} is a mechanism which competes with the usual GUT baryogenesis channel based on the out-of-equilbrium decay of the superheavy $A_{\mu}$ and $X$ particles. If $T_{out} \ll T_{GUT}$, then defect-mediated GUT baryogenesis is in fact the dominant mechanism.

\section{Dynamical Breaking of CPT and Defect-Mediated Baryogenesis}

Let us in the following consider explicitly how dynamical CPT violation is crucial for defect-mediated baryogenesis \cite{PTDB}. To be specific, we consider extensions of the standard model with CP violation in an extended Higgs sectior in the form of a CP violating phase $\epsilon$ (e.g. the relative phase between the two doublets in the two Higgs doublet model). This phase has the following transformation properties under CP and T:
\begin{eqnarray}
CP: & \,\,\, & \epsilon(x,t) \, \rightarrow \, - \epsilon(-x,t) \nonumber \\
T: & \,\,\, & \epsilon(x,t) \, \rightarrow \, - \epsilon(x,-t) \\
CPT: & \,\,\, & \epsilon(x,t) \, \rightarrow \, \epsilon(-x,-t) \, .\nonumber
\end{eqnarray}
Hence, $\partial_{\mu} \epsilon$ is odd under CPT.

How the defect (which we take to be a string) interacts with the plasma can be modelled by a term in the Lagrangian of the form
\begin{equation}
{\cal L}_{\epsilon} \, \propto \, \partial_{\mu} \epsilon j_5^{\mu} \, ,
\end{equation}
where $j_5^{\mu}$ is the axial current \cite{JPT}. The axial current transforms under CPT as
\begin{equation}
CPT: \,\,\,\, j_5^{\mu} \, \rightarrow \, -j_5^{\mu} \, ,
\end{equation}
so that the interaction Lagrangian ${\cal L}_{\epsilon}$ is invariant under CPT as it must be. Hence, it follows that a static string is its own antiparticle, and hence cannot generate any net baryon number. 

For a moving string, an apparent CPT paradox arises: The CPT conjugate of a defect with a value $\epsilon > 0$ inside the core moving with velocity ${\vec v}$ is a defect with the same value of $\epsilon$ in the core moving with the same velocity vector ${\vec v}$. Hence, if CPT were a symmetry, then it would not be possible for the string to generate a net baryon number.

The resolution of this apparent CPT paradox starts with the observation that the master equation (\ref{master}) for baryogenesis is a dissipative equation which explicitly violates T symmetry, and, since it conserves CP, also violates CPT. Like the Ilion field of Cohen and Kaplan \cite{CK}, the defect network evolution drives the system out of thermal equilibrium, acting as an external source of T violation, and the dissipative processes tend to restore the thermal equilibrium. In turn, dissipation leads to a damping of the defect motion. If dissipation were the only force, then the defects would come to rest and $n_B$ violation would stop. However, the expansion of the Universe induces a counterforce on the defects which keeps the defect network out of equilibrium and allows $n_B$ generation to continue.

The lesson we draw from this study is that the expansion of the Universe is the source of explicit external T violation which keeps the defect network out of equilibirium. The ordering dynamics of the defect network fueled by the cosmological expansion then leads to dynamical CPT violation and to the biasing of baryon number production.

\section{Defect-Mediated QCD Scale Baryogenesis}

As mentioned above, defect-mediated baryogenesis can be effective not only at the electroweak scale, but at any scale when defects are produced. Recent work \cite{HZ} has shown that as a consequence of the nontrivial vacuum structure of low energy QCD, domain walls form at the QCD phase transition. These domain walls separate regions of space in which the effective strong CP parameter $\theta$ has very different values. Hence, the domain walls automatically are associated with maximal CP violation. 

Recently\cite{BHZ}, a new baryogenesis (more precisely charge separation) scenario was proposed based on these QCD domain walls. Since this mechanism will be reviewed in a separate conference proceedings article \cite{BHZ2}, I will here only highlight the main points.

The starting point of the QCD baryogenesis scenario is a new nonperturbative analysis of the vacuum structure of low energy QCD \cite{HZ}. Considering the vacuum energy $E$ of pure gluodynamics as a function of $\theta$, it was realized \cite{HZ} (see also \cite{Shifman,Witten}) that $E(\theta)$ must have a multi-branch structure
\begin{equation}
E(\theta) \, = \, {\rm min}_k E_k(\theta) \, = \, {\rm min}_k E_0(\theta + 2 \pi k) \, ,
\end{equation}
and hence must in general have several isolated degenerate minima.

When fermionic matter is introduced, at low energies represented by a chiral condensate matrix $U$ which contains the pion and sigma prime fields, then the potential energy $W(U, \theta)$ depends only on the combination $\theta - i Tr {\rm ln} U$ (by the anomalous Ward Identities). Hence, from the multi-branch structure of $E(\theta)$ it immediately follows that for fixed value of $\theta$, the potential $V(U) = W(U, \theta)$ has several isolated minima. These vacua differ in terms of the effective strong CP parameter $\theta$. 

Since there are several discrete minima of the potential, domain walls separating the different phases exist. In fact, by the Kibble mechanism \cite{Kibble}, during the QCD phase transition at $T = T_{QCD}$, inevitably a network of domain walls will form.

The second crucial ingredient of the new scenario \cite{BHZ} is charge separation. In analogy to how in $1+1$ dimensional physics solitons acquire a fractional charge \cite{JR}, in a $3+1$ dimensional theory domain walls will also acquire a fractional baryonic charge.

In the chiral limit, the different vacuum states would be energetically degenerate. In the presence of a nonvanishing quark mass $m_q$, the energy of states increases as a function of $|\theta|$. Hence, the different phases of the theory, which immediately after the phase transition are equidistributed, will no longer be so below a temperature $T_d$ at which the energy difference between the minima becomes thermodynamically important. At this time, the domain wall network will break up into a set of {\it B-shells}, domains of states of large $|\theta|$ in a surrounding sea of the phase with the lowest value of $|\theta|$. In the absence of explicit strong CP violation, i.e. when the lowest energy vacuum is $\theta = 0$, then there are an equal number of B-shells with
$\theta > 0$ and $\theta < 0$. A B-shell with $\theta > 0$ has \cite{BHZ} negative baryon number. 

In order to generate a net baryon number in B-shells, (a small amount of) explicit CP violation is required such that the only B-shells which form have the same sign of $\theta$ (which we take to be positive). In this case, the total baryon number trapped in the walls is negative. Since there is no overall baryon number violation in QCD, the compensating positive baryon number must be in the bulk. This is the way in which domain walls in QCD lead to an effective baryogenesis mechanism by means of charge separation. Note that, in analogy to electroweak baryogenesis, the explicit T violation due to the expansion of the Universe which leads to the coarsening and eventual fragmentation of the defect network is crucial for the mechanism. As our estimates \cite{BHZ} indicate, it appears possible to generate a baryon to entropy ratio comparable to what observations require.

\section*{Acknowledgments}

I wish to thank Alan Kostelecky for the invitation to speak at this meeting, and for his hospitality. I thank my collaborators Anne-Christine Davis, Igor Halperin, Tomislav Prokopec, Mark Trodden and Erik Zhitnitsky, for enjoyable and stimulating collaborations. This work was supported in part by the US Department of Energy under Contract DE-FG02-91ER40688.

\end{document}